# FROM USER-UNDERSTANDABLE TO TECHNICAL PROCESS MODEL: A MODEL-DRIVEN APPROACH USING CUTA4BPM

Kathrin Kirchner[1], Siniša Nešković[2], Dejan Stojimirović[2]
***[1]University Hospital Jena, Erlanger Allee 101, Jena, Germany***
***[2]Branislav Lazarević Laboratory of Information Systems***
***Faculty of Organizational Sciences, University of Belgrade, Jove Ilića 154, Belgrade, Serbia***

**Abstract –** *For business process modeling, we can choose between graph-oriented and block-oriented languages. Block-oriented languages are more structured and therefore better understandable for domain experts, while graph-oriented languages allow more modeling freedom and technical versatility for process designers. To bridge this gap between understandability and technical versatility, we propose a participative forward engineering approach. It uses our block-oriented CUTA4BPM language to support high level process modeling together with domain experts, while graph-oriented BPMN is used for further detailed process design and automation. To support smooth transition from the high level modeling step to the detailed one, CUTA4BPM process models are automatically transformed to BPMN models using model driven engineering techniques.*

## 1. INTRODUCTION

Participative techniques [16] for process modeling, e.g. CUTA card game [11], aim at modeling processes in a cooperative session. Domain experts and process analysts sit together and use paper cards to discuss and model the process together. The drawback of such a method is the resulting informal specification of the model. Answering this drawback, several approaches exist to make formal process models fit for domain experts, e.g. t.BPM [2]. Based on BPMN notation [13], it uses tangible media in the form of plastic shapes that can be labeled using a pen. The shapes can be laid on a table, where they can be connected with lines. Although this approach supports cooperative modeling, it is still not so easy for domain experts to understand and deal with the graph-oriented nature of t.BPM and BPMN. Most importantly, this approach does not prevent domain experts to create invalid process models (e.g. to make a deadlock).

Regarding business process modeling languages, we can distinguish between graph-oriented and block-oriented ones [10]. The main advantage of block-oriented languages, comparing to graph oriented ones such as BPMN, is that they are significantly less error prone and safer. Additionally, due to the block structure, they require less number of symbols and, hence, fewer control flow elements. Therefore they are easier for understanding and usage by non-IT experts compared to graph-oriented languages. Consequently, block-oriented languages provide better support for participative design between domain experts and IT experts. On the other hand, graph oriented languages, such as BPMN, are more suitable for further technically detailed process design and process automation.

In order to overcome these conflicting requirements and bridge this gap between expert understandability and technical versatility, we propose a new participative forward engineering approach which is based on model driven engineering techniques. It uses our special block-oriented CUTA4BPM language [9] to support high level process modeling together with domain experts, while standardized graph-oriented BPMN is used for further detailed process design and automation. To support smooth transition from the high level modeling step to the detailed one, CUTA4BPM process models are automatically transformed to BPMN models using an automatic model transformation.

The paper is structured as follows. First, we describe existing modeling approaches and discuss their suitability of eliciting and formalizing expert-driven processes. In section 3, we introduce the modeling notation CUTA4BPM card game and our participative forward engineering approach. Section 4 explains the model transformation from CUTA4BPM to the standard modeling language BPMN. Section 5 provides the conclusions of our work.

## 2. RELATED WORK

A number of graphical modeling languages have been developed to support analysis and modeling of processes in companies. Graph-oriented languages, like BPMN, UML AD [14] or EPC [7], specify control flow using arcs that define the temporal and logical dependencies between nodes. Different types of nodes can exist in such a language, whose number sometimes can be rather large, as it is the case with BPMN. Additionally, only few restrictions exist for graph-oriented languages, since edges can connect almost arbitrary types of nodes.

In block-oriented languages, control flow is defined by nesting control primitives that are used to represent concurrency, alternatives or loops [10]. The idea can be compared with typical imperative programming languages. Contrary to graph-oriented languages, that can define arbitrary control flows, block-oriented languages only allow to model structured control flows.

While graph-oriented and unstructured models provide more freedom and versatility to IT experts, it is easier to define a valid workflow using block-oriented structured workflows, e.g. to make deadlock free process models [10]. Holl and Valentin discuss the advantages of a structured business process modeling, e.g., models will become more transparent. They can be more easily documented, adapted and modified in a graphical or verbal way [5].

Transforming a block-oriented model to a graph oriented one is done using model-driven engineering (MDE), a software development approach which relies on models as main artifacts [1]. In MDE, several types of models are used, each of them representing a different abstraction level. The first one is the computation-independent model (CIM) which is focused on environment and system requirements independently of the system that has to be developed. On the next abstraction level, the platform-independent model (PIM) concentrates on the parts of the system that will be computerized. Still, this PIM abstracts from the technical platform of implementation. The next layer is called platform-specific model (PSM), which describes the specific characteristics of the used platform.

Transformations from one model to another model are important steps in MDE. A transformation is defined as automatic generation of a target model from a source model according to transformation rules. Transformation rules, specified using some transformation language, describes how one or many constructs from the source language can be transformed into one or more constructs from the target language. In this work, QVT operational language [15], which is an OMG standard, is used as a transformation language.

When transforming a block-oriented model to a graph-oriented one, a nested block structure has to be mapped on a graph, and one block can result in a whole subgraph. Mendling et al. [12] propose three strategies for the transformation from block-oriented to graph-oriented languages. One of them is the flattening strategy, that transforms a nested control flow into a at process graph without hierarchies. The nested structure is therefore traversed recursively. This approach is used in our work too.

## 3. CUTA4BPM AND PARTICIPATIVE FORWARD ENGINEERING APPROACH

For eliciting expert-driven processes together with domain experts, a suitable modeling language was needed that allows a precise specification of processes understandable for non-IT experts. With CUTA4BPM card game [9], the knowledge of the process expert as well as domain-specific terminology can be elicited and formalized in a cooperative way between expert and process analyst.

The basic element of CUTA4BPM is the SimpleActivity card (shown in the Fig. 1).

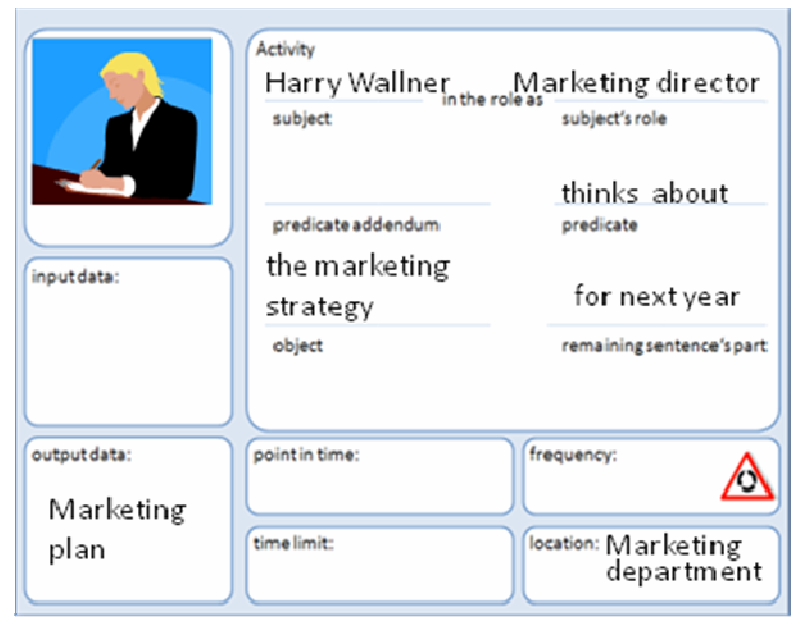

**Figure 1. SimpleActivity card in CUTA4BPM**

The activity description is a complete sentence in natural language (subject – predicate - object) which provides the content related information for the corresponding activity in the business process. The subject is normally a concrete person. Additionally, the subject's role describes whether this activity could be performed by a business unit or a group of people where the subject belongs to. Furthermore, it can be described which *documents* are needed to conduct the activity, where it takes place (*location*) or whether there is any *time limit.*

Furthermore, control flow blocks are needed to describe a process. They are defined in Table 1.

| Control flow block | Description |
|---|---|
| Sequence | Sequence elements like Simple Activities or other blocks executed one after the other |
| Case | [condition] [condition] ... Execute one out of many alternatives |
| Loop | [condition] [condition] Repeat a sequence of elements |
| Parallel | Do All! Constraints: Execute at least two elements in parallel |
| Multiple choice | Do All! Constraints: [constraint] One or more out of many elements are executed |

**Table 1. Control flow blocks in CUTA4BPM**

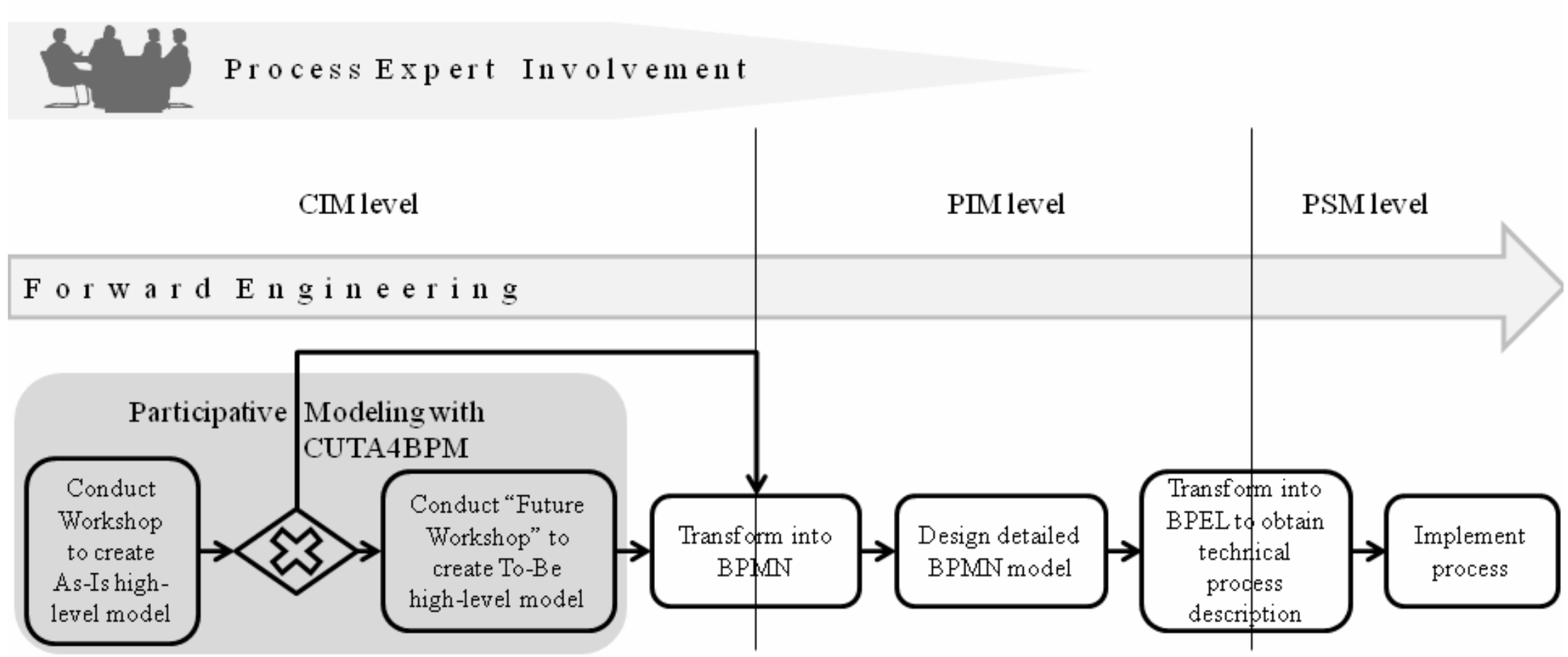

**Figure 2: Participative forward engineering approach**

CUTA4BPM is aimed to support just a part of the business process management life cycle, covering the CIM level of analysis and conceptual processes modeling. It is not intended to support detailed design and, hence, is not suitable as a basis for automation of processes. That is why it needs to be translated to more formal oriented languages.

In this work, MDE development is fully supported with CUTA4BPM serving as a computer independent model (CIM), while BPMN and BPEL serve as a platform independent model (PIM) and a platform specific model (PSM) respectively. Figure 2 shows our participative model-driven forward engineering lifecycle for business process management.

BPM lifecycle starts with knowledge elicitation and process analysis resulting in a CUTA4BPM model. The cards are applied in interviews between process analyst and process expert. Each interview consists of two sessions which are conducted on two different days. The first session aims to obtain an overview of main characteristics of a business process. In the second session, the process analyst summarizes the results of the first session with the help of the cards. In a participative way, cards can be rearranged or added to result in a more precise high-level process model.

The elicited as-is-process can be discussed with the experts to find optimization potential and develop a to-be-process. To develop ideas for future changes the idea of future workshops [6] is used which allows an active participation of all stakeholders. Thus, process experts can discuss critical drawbacks and possible changes of the current process using its visualization in CUTA4BPM.

The resulting CUTA4BPM model is then transformed into a BPMN model, which is later used for detailed process specification and design. The resulting BPMN model can be extended to result in a more detailed model, enabling later transformation to BPEL used for process execution.

Although BPMN and other graph oriented languages are perhaps better equipped for specification of all technical aspects of processes, especially for specification of process automation, this is not an issue in participative design and such detailed specifications are postponed for later phases of process development. In this way, separating high level process specification from its low level (technical) design, complexity of process development is better managed.

## 4. TRANSFORMATION FROM CUTA4BPM TO BPMN

According to the MDE approach, model transformations require the existence of meta models for both source and target models. BPMN meta model is a part of the OMG standard [13] and it will not be described here. CUTA4BPM metamodel is shown in Fig. 3 and described in the following.

Blocks can be Sequence, Case, Parallel, Loop or MultipleChoice. A sequence consists of sequence elements which are numbered by SeqNo. A case block consists of case items that are described by a condition and a number. For a loop, a loop condition and its position (beginning or end of loop) has to be defined. For multiple choice, a condition is defined. A SimpleActivity has several attributes as they are defined on a CUTA4BPM activity card. Input and output documents, roles and organizational units can occur on several activities of a CUTA process. Therefore they are described separately in extra classes in the metamodel. A role is part of an organizational unit.

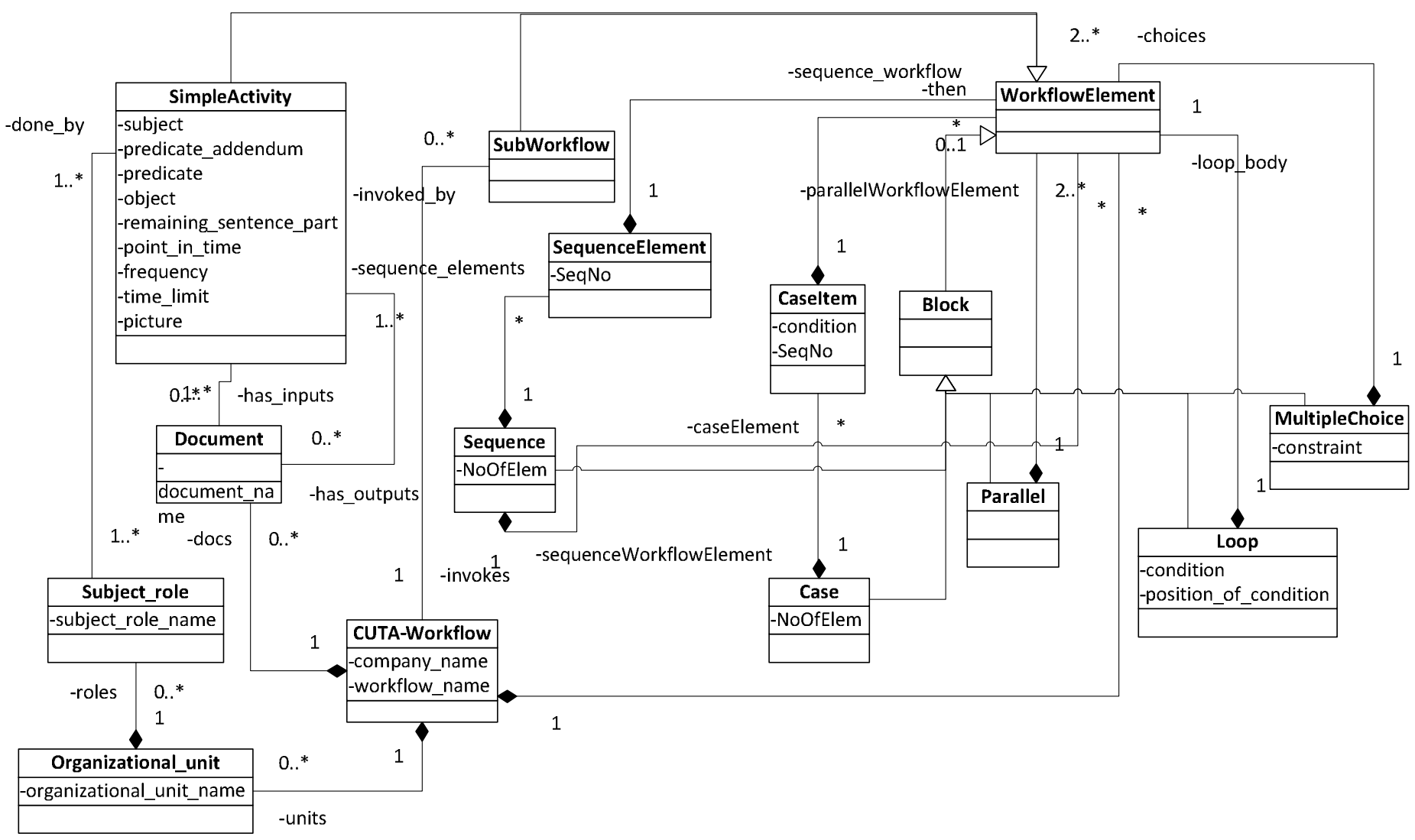

**Figure 3: CUTA4BPM metamodel**

The same document can be input (or output) of one activity and also input (or output) of another activity.

Based on the meta models of CUTA4BPM and BPMN, a model transformation using QVT operational language is implemented. The main issue in the transformation of CUTA4BPM to BPMN is that CUTA4BPM is defined as a block language, while BPMN is a graph-oriented language. That means that a nested block structure has to be mapped on a graph, and one block can result in a whole subgraph. In our work we used the flattening pattern approach [12], which is illustrated in Fig. 4.

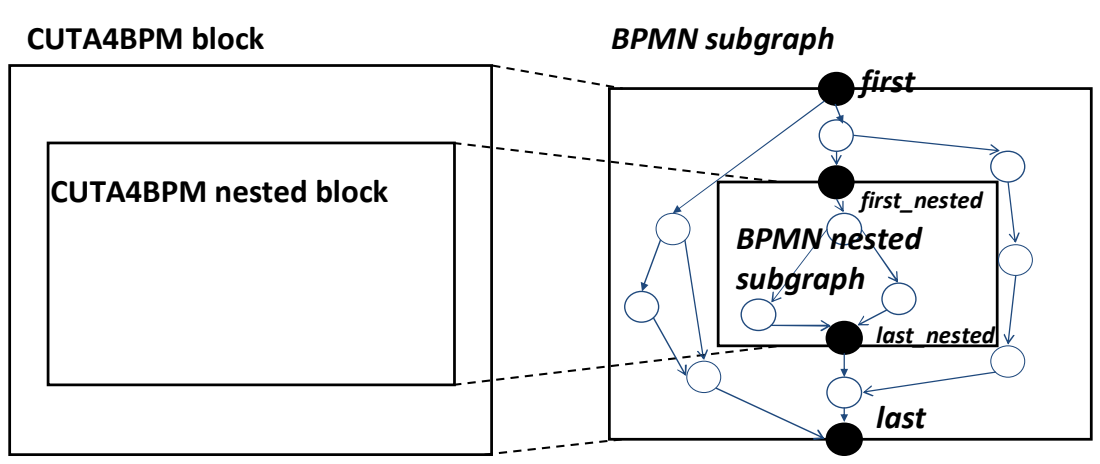

**Figure 4: The flattening strategy**

During transformation, every CUTA4BPM block is mapped to a BPMN subgraph. Each subgraph has a first and last element representing entry and exit points of the subgraph. Nested blocks are translated as subgraph of another subgraph recursively applying the adopted flattening strategy. The exact shape of mapped subgraphs depends on the type of CUTA4BPM nested block mappings. In other words, each type of CUTA4BPM block maps to a corresponding BPMN subgraph pattern. Fig.5 shows an example for the mapping with the CUTA4BPM workflow.

### 4.1. Mapping CUTA4BPM Workflow

CUTA4BPM workflow is directly mapped to a BPMN process diagram, consisting of the start and end events, which represent the top most starting and ending nodes in the whole BPMN graph, as it is illustrated in the following figure.

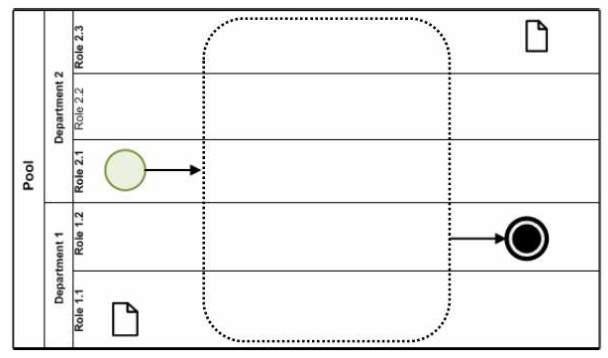

**Figure 5: Transformation of CUTA4BPM workflow**

Additionally, a corresponding pool is also created representing the company the modeled process belongs to.

### 4.2. Mapping SimpleActivity to BPMN Task

SimpleActivity block is mapped to a BPMN task, consisting of corresponding input and output documents (Fig 6). Furthermore, an additional lane is created for the corresponding role performing the activity (task), if it is not already existing in the BPMN graph.

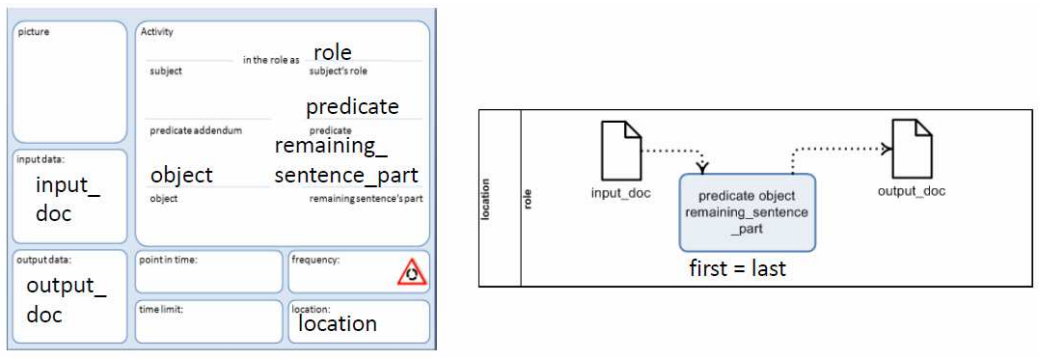

**Figure 6: Transformation of SimpleActivity**

### 4.3. Mapping from Sequence block to BPMN

A CUTA4BPM sequence consists of an ordered array of blocks or simple activities. It is transformed to a sequence of BPMN subgraphs that are connected with sequence flows (arrows), as it s shown in Fig. 7.

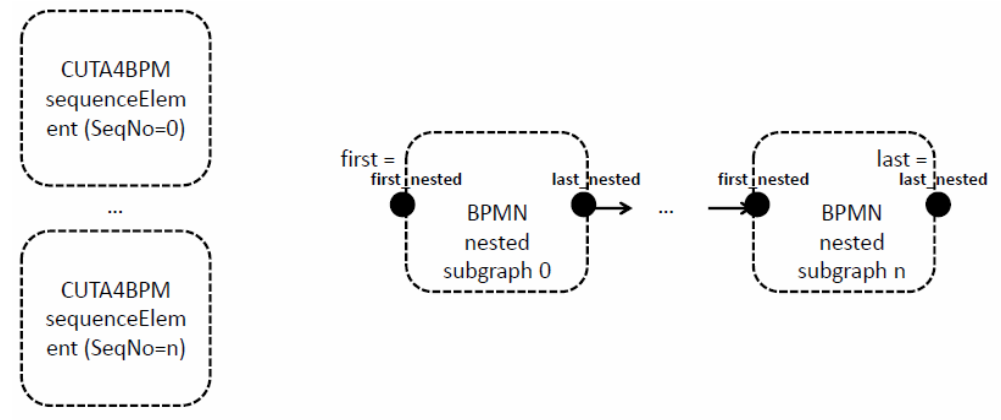

**Figure 7: Transformation of case block**

### 4.4. Mapping case block

A case block generally results in a BPMN subgraph starting and ending with exclusive gateways (Fig. 8). Therefore, a fork exclusive gateway is created as first element of BPMN subgraph, while the last element of BPMN subgraph is a join exclusive gateway.

.

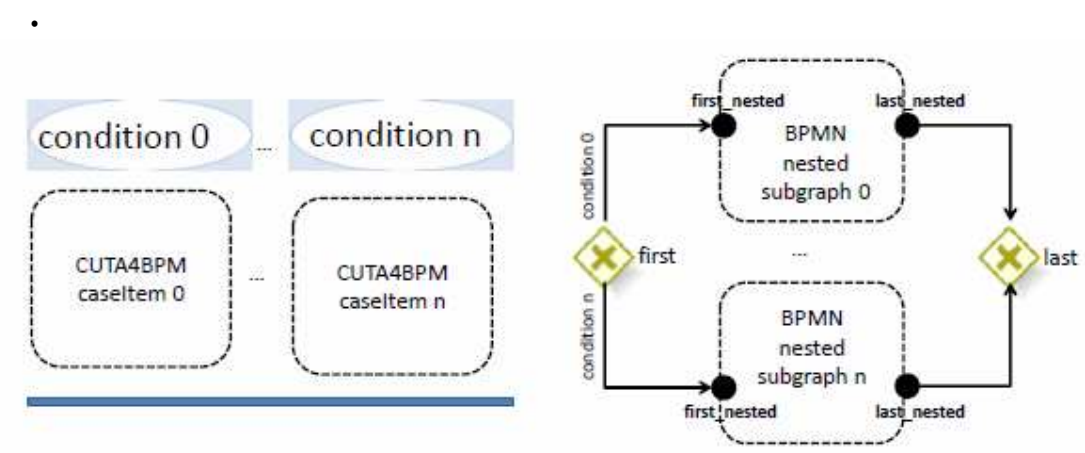

**Figure 8: Transformation of case block**

The condition that is written on a case item is translated to the condition expression of the corresponding sequence flow

### 4.5. Mapping parallel block

For expressing a parallel workflow, parallel gateways are used in BPMN (Fig. 9).

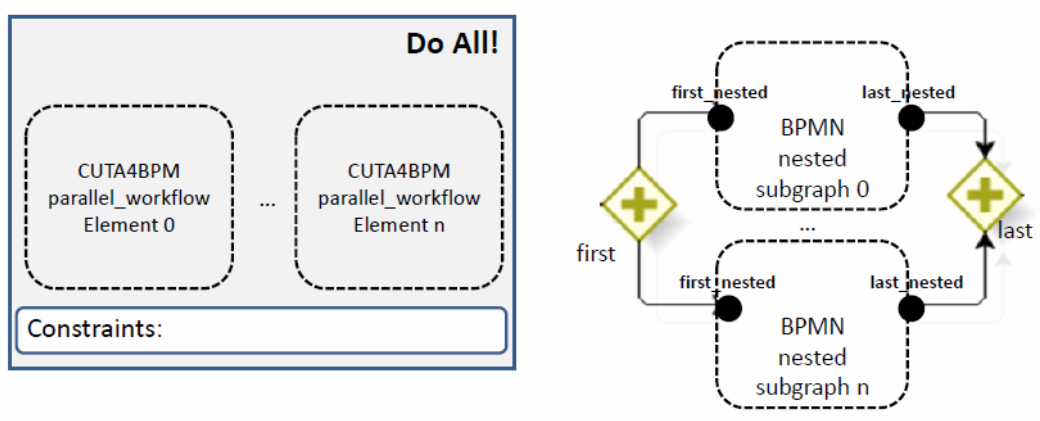

**Figure 9: Transformation of parallel block**

A fork and join parallel gateway are created as first and last element of the corresponding BPMN subgraph.

### 4.6. Mapping multiple choice block

The multiple choice block looks in CUTA4BPM quit similar to the parallel block, but additional has a constraint. Here, the inclusive gateway is used in BPMN (Fig. 10).

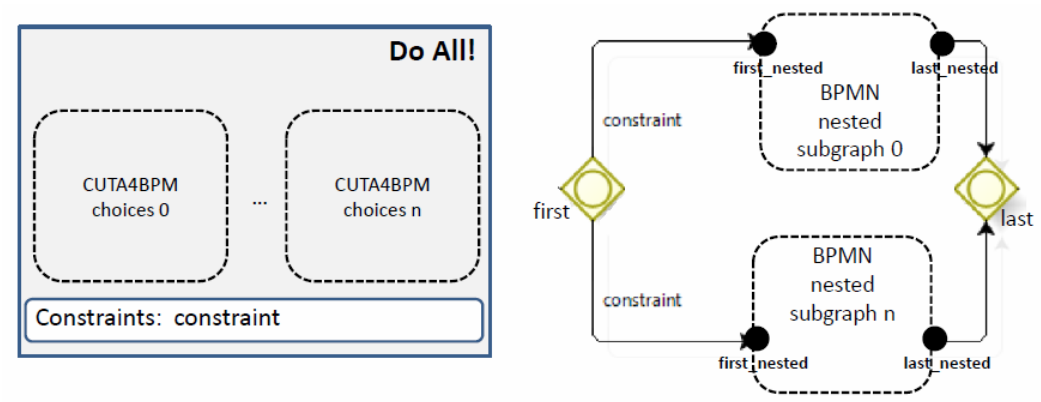

**Figure 10: Transformation of multiple choice block**

The mapping is done in a similar way like the parallel block translation (but using inclusive instead of parallel gateway).

### 4.6. Mapping loop block

In case of case block, the corresponding BPMN subgraph has an exclusive gateway as first and last element and every block in the loop body has to be mapped to a corresponding BPMN subgraph. The exact pattern for transformation of the loop block depends on the position of the loop condition. If the condition is at the beginning of the loop (Fig. 11), a direct sequence flow is created between the first element of BPMN subgraph (the gateway) and the next gateway.

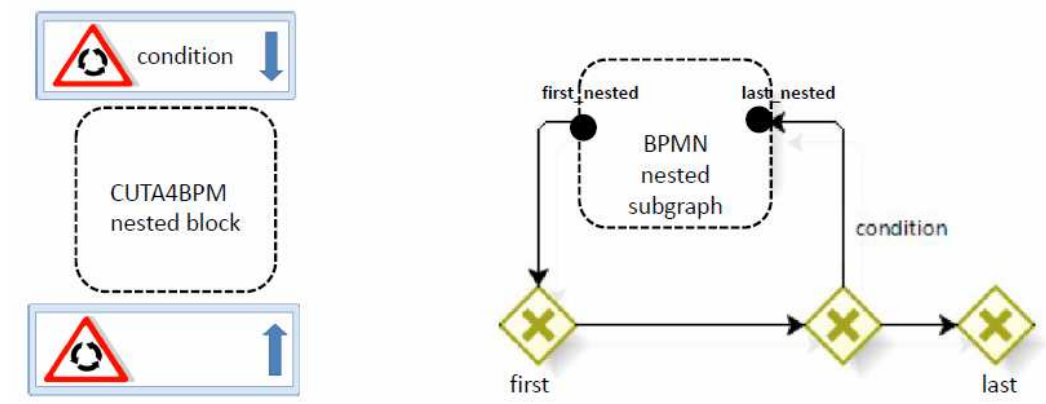

**Figure 11: Transformation of loop block**

In case the loop condition is at the end of the loop (Fig. 12), a BPMN sequence flow connects the last element of the nested BPMN subgraph with the second exclusive gateway.

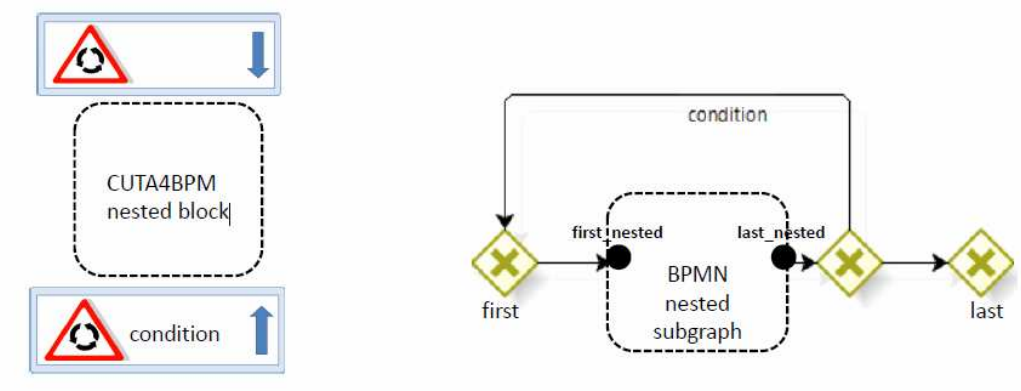

**Figure 12: Transformation of loop block**

## 5. CONCLUSION

In this paper we introduced a MDE approach of a participative business process lifecycle based on CUTA4BPM language. CUTA4BPM is used for analysis and modeling of business processes together with the domain experts. The obtained CUTA4BPM process model can be transformed into the more formal BPMN and BPEL models, which provides a basis for a more detailed technical modeling as well as for an implementation. Also, a back transformation from BPEL and BPMN models into CUTA4BPMN is possible. Thus, both forward and backward (re)engineering of a process is supported.

We verified our CUTA4BPM cards in nearly 20 interviews with process experts working in farming [8], pharmacy, university administration [3] and production. [4]. As our evaluation with experts from various domains shows, our approach helps to model processes driven by expert knowledge and expertise together with the process expert. Structuring of thoughts is easier for experts, modeling by examples is supported, communication is improved and finally the risks of imprecise and incomplete process models are reduced.

Regarding software support for this approach, a prototype of the CUTA4BPM graphical editor as well as a transformation into BPMN are already implemented. Next, we plan to support backward engineering.

## ACKNOWLEDGEMENT

Research presented in this paper was partially supported by the Government of Republic of Serbia, Project Grant III-44010, Title: Intelligent Systems for Software Product Development and Business Support based on Models

## REFERENCES

[1] Brambilla, M., Cabot, J., & Wimmer, M. (2012). *Model-Driven Software Engineering in Practice*. Morgan & Claypool Publishers.

[2] Edelman, J., Grosskopf, A., Weske, M., & Leifer, L. (2009). Tangible business process modeling: A new approach. In Proc. of 17th Intern. Conf. on Engineering Design.

[3] Erfurth, I. & Kirchner, K. (2010). *Requirements Elicitation with Adapted CUTA Cards: First Experiences with Business Process Analysis*. In 5th IEEE Intern. Conf. on Engineering of Complex Computer Systems (pp. 215-223).

[4] Günther, G. (2011). *Das Kartenspiel C4U - Eine Evaluation aus der Perspektive des Entwicklers und des Kunden*. Bachelor Thesis, Friedrich Schiller University Jena, Germany.

[5] Holl, A. & Valentin, G. (2004). *Structured Business process modeling (SBPM)*. In Inform. Systems Research in Scandinavia (IRIS 27).

[6] Jungk, R. & Mullert, N. (1996). *Future Workshops: How to Create Desirable Futures*. Institute for Social Inventions, 2nd edition.

[7] Keller, G., Nüttgens, M., & Scheer, A.-W. (1992). *Semantische Prozessmodellierung auf der Grundlage Ereignisgesteuerter Prozessketten (EPK)*. In A.-W. Scheer (Ed.), Veröffentlichungen des Instituts für Wirtschaftsinformatik, 89, Saarbrücken.

[8] Kirchner, K., Erfurth, I., Möckel, S., Gläßer, T., Schmidt, A. (2010): *A Participatory Approach for Analyzing and Modeling Decision Processes: A Case Study on Cultivation Planning*, In: Manos, B. et al. (Eds.): Decision Support Systems in Agriculture, Food and the Environment: Trends, Applications and Advances, (pp. 138-154).

[9] Kirchner, K. & Nešković, S. (2012). *Using CUTA4BPM to Support Participative Development of Expert-Driven Processes*. In Proc. of ICIST 2012 - 2nd International Conference on Inform. Society Technology (pp. 41-46).

[10] Kopp, O., Martin, D., Wutke, D., Leymann, F.: *The Difference Between Graph-Based and Block-Structured Business Process Modelling Language*, *Enterprise Modelling and Information Systems Architectures*, Volume 4, Issue 1, 2009, (pp. 3-13).

[11] Lafreniére, D. (1996). *CUTA: A simple, practical, low-cost approach to task analysis*. Interactions, 3(5) (pp. 5-39).

[12] Mendling, J., Lassen, K. B., & Zdun, U. (2008). *On the transformation of control flow between block-oriented and graph-oriented process modelling languages*. Int. J. of Business Process Integration and Management, 3(2), 96-108.

[13] OMG (2010). *Business Process Model and Notation (BPMN) Version 2.0*. www.omg.org/cgi-bin/doc?dtc/10-06-04.pdf

[14] OMG (2010). *OMG Unified Modeling Language Superstructure Version 2.3*. www.omg.org/spec/UML/2.3/Superstructure

[15] OMG (2011). *Meta Object Facility (MOF) 2.0 Query/View/Transformation Specification*. www.omg.org/spec/QVT/1.1/PDF/

[16] Schuler, D., Namioka, A.: *Participatory design*. Hillsdale, NJ: CRC / Lawrence Erlbaum Associates, 1993.